\documentclass[a4paper,12pt]{article}
\usepackage[cmex10]{amsmath}
\usepackage{latexsym,amssymb}
\usepackage[mathscr]{eucal}
\usepackage{amsxtra,amscd,amsthm,upref,layout,color}
\usepackage{calc}
\usepackage[final]{graphicx}
\def\refup#1{{$^{#1}$}}
\def\p{^{^{\prime}}}\def\pp{^{^{\prime\prime}}}

\def\oo{{\leavevmode\setbox0=\hbox{h}\dimen0=\ht0 \advance\dimen0
by-1ex\rlap{\raise0.47\dimen0\hbox{\char'27}}o}}
\def\begeq{\begin{equation}}
\def\endeq{\end{equation}}
\def\begdis{\begin{displaymath}}
\def\enddis{\end{displaymath}}

\def\cT{{\cal T}}

\def\arctan{\rm{arctan}}\def\arcos{\rm{arcos}}
\def\alft{\alpha_{_{\rm T}}}

\def\ie{{\em i.e.}}\def\eg{{\em e.g.}}
\def\etal{{\em et al.}}

\begin{document}
\title{The correlation functions  of the regular tetrahedron and octahedron}
\author{  
{{Salvino Ciccariello}}\\
  \begin{minipage}[t]{0.9\textwidth}
   \begin{flushleft}
   \setlength{\baselineskip}{12pt}
{\slshape  {\footnotesize{Universit\`{a} di Padova,
   Dipartimento di Fisica {\em G. Galilei}}}}\\
  {\slshape{\footnotesize{Via Marzolo 8, I-35131 Padova, Italy}} }\\
  \footnotesize{salvino.ciccariello@unipd.it}
\end{flushleft}
\end{minipage}
}
\date{July 10, 2014}
\maketitle                        
\begin{abstract} \noindent
The expressions of the autocorrelation functions (CF) of the regular tetrahedron
and the regular octahedron are reported. They have an algebraic form
that involves the arctangent  function and rational functions of $r$
and  $(a+b\,r^2)^{1/2}$,  $a$ and $b$ being
appropriate integers and $r$ the distance.  The CF expressions make the
numerical determination of the corresponding scattering intensities
fast and accurate also in presence of a size dispersion. \\  \\
\noindent Synopsis: {\em  One reports the algebraic expressions of the tetrahedron's and
octahedrons' correlations functions.}\\

\end{abstract}
\vfill
\eject
{}{}
\subsubsection*{1  Introduction}
The algebraic knowledge of the autocorrelation function (CF) of a homogeneous particle,
with a given shape,  is useful from an applied point of view because
polydisperse analyses with such a geometrical shape no longer require a numerical
integration to derive the particle form factor.  Of course
few  geometrical shapes have  algebraic CFs. Up to now  it has
been shown that  the CFs of the sphere (Guinier \& Fournet, 1955),
the cube (Goodisman, 1980) and the right parallelepiped (Gille, 1999)  only
have an algebraic form. Those of the rotational cylinder (Gille,1987 )
and ellipsoid (Burger \& Ruland, 2001) do not have this property since they
involve non-elementary transcendental functions, as the elliptic ones.
For a recent review, see Gille (2013).
Recently it has been reported that  two more Platonic
solids have an algebraic chord-length probability density, \ie\ the second
derivatives [$\gamma\pp(r)$]  of the relevant CFs [$\gamma(r)$] have an
algebraic form. The two solids are the regular tetrahedron (Ciccariello, 2005a)
and the regular octahedron (Ciccariello, 2014).
We have realized that these two second order derivatives can explicitly be
twice integrated.\\
This short paper is devoted to  report the resulting algebraic expressions,
which are also of  interest both in Rietveld analysis of wide angle diffraction patterns
and in the  stochastic geometry realm (Chiu \etal\ 2013) because particle isotropic
covariograms are simply related to  particle CFs.
\subsubsection*{2  CF expressions}
To explain the  derivation of these CFs it is sufficient to note that the key
ingredient are: {\em i)} the  integration by parts of each term contributing to
$\gamma\pp(r)$ and having  the form $f(r)\cT\left(h(r)\right)$ ($\cT$ being
an inverse trigonometrical function), and {\em ii)} a convenient regrouping of
the resulting  terms. In fact,for some of these terms,  after  an integration by parts 
the resulting integrand
$g(r)\,h\p(r)\,\cT\p\left(h(r)\right)$ [where $g(r)$ denotes a primitive of $f(r)$]
has the form $R\big(r,P_2(r)^{1/2}\big)$,
where $R(r,y)$ is a rational function and $P_2(x)$ a 2nd degree polynomial.
Thus, according to a general mathematical result [see, \eg, Caccioppoli,
1955], the primitive of $g(r)\,h\p(r)\,\cT\p\left(h(r)\right)$ has a simple
algebraic form that is simply and safely worked out by MATHEMATICA.
(The MATHEMATICA codes used to get the following results are found at
{\em http://digilander.libero.it/sciccariello/}.)
In the other cases, say $\cT_1(r)$,..,$\cT_4(r)$, where $g_i(r)\,\cT_i\p(r)$  has not the
mentioned form, the mentioned property holds true after performing a suitable regrouping,
say $g_1(r)\,\cT_1\p(r)+g_3(r)\,\cT_3\p(r)$ and
$ g_2(r)\,\cT_2\p(r)+g_4(r)\,\cT_4\p(r)$.\\
It is now convenient to put
\begeq                     
S_{_{\rm T}}\equiv(3)^{1/2},\quad V_{_{\rm T}}\equiv 1/[6 (2)^{1/2}],
\quad\alpha_{_{\rm T}}\equiv\arcos(1/3),
\endeq
\begeq
S_{_{\rm O}}\equiv2\,({3})^{1/2},\quad V_{_{\rm O}}\equiv ({2})^{1/2}/3,
\quad\alpha_{_{\rm O}}\equiv\arcos(-1/3),
\endeq
where indices T and O respectively refer to the tetrahedron and the octahedron
of unit sides,
and
\begeq
\Delta_{34}(r)(r)\equiv(4\,r^2-3)^{1/2},\quad \Delta_{11}(r)(r)
\equiv (r^2-1)^{1/2},
\endeq
\begeq
\cT_1(r)\equiv\frac{4 r^4-12 r^2+7}{4\left(1-r^2\right)\Delta_{34}(r)},
\quad
\cT_2(r)\equiv  \frac{9 r^2-7}{3 (3)^{1/2}\, (1-r^2)\Delta_{34}(r)},
\endeq
\begeq
\cT_3(r)\equiv\frac{2 (2)^{1/2}\, r+3}{\Delta_{34}(r)},\quad
\cT_4(r)\equiv \frac{(2)^{1/2}\, r}{\Delta_{34}(r)},\quad
\cT_5(r)\equiv \frac{\Delta_{34}(r)}{2\,(1- r^2)},
\endeq
\begeq
\quad \cT_6(r)\equiv \frac{\Delta_{34}(r)}{(3)^{1/2}},\quad
\cT_{7}(r)\equiv \frac{6\,r^2-5}{(3)^{1/2}\,\Delta_{34}(r)},\quad
\cT_{8}(r)\equiv \frac{r\,\left(6\,r^2-5\right)}{(2)^{1/2}\,
\Delta_{34}(r)},
\endeq
\begeq
\cT_{9}(r)\equiv \frac{2 r^4+r^2-2}{2 (2)^{1/2}\, r\,(1-r^2)\,
\Delta_{34}(r)},\quad
\cT_{10}(r)\equiv \frac{(3)^{1/2}\,\Delta_{34}(r)}{2\,r^2-3},
\endeq
\begeq
\cT_{11}(r)\equiv\frac{\Delta_{34}(r)+1}{\Delta_{34}(r)-1},\quad
\cT_{12}(r)\equiv \frac{2 r^2-3}{(3)^{1/2}\, \Delta_{34}(r)},\quad
\cT_{13}(r)\equiv\frac{-2 r^4+12 r^2-9}{(3)^{1/2}\, (2r^2-3)\,
\Delta_{34}(r)},
\endeq
\begeq
\cT_{14}(r)\equiv\frac{(3)^{1/2} (2\, r^8-34\, r^6+
  96\,r^4-90\, r^2+27)}{(10\, r^6-54\, r^4+72\,r^2-27)
  \Delta_{34}(r)},\quad \cT_{15}(r)\equiv (3)^{1/2}\,\Delta_{11}(r) ,
\endeq\begeq
\cT_{16}(r)\equiv \frac{r}{(2)^{1/2}\,\Delta_{11}(r)},\quad
\cT_{17}(r)\equiv\frac{7\, r^4-4\, r^2-4}{4\, (2)^{1/2}\, r\,
\left(r^2-2\right)\, \Delta _{11}(r)},
\endeq
\begeq
\cT_{18}(r)\equiv\frac{9 r^2-10}{3 (3)^{1/2}\left(r^2-2\right)
\Delta _{11}(r)},\quad
\cT_{19}(r)\equiv\frac{r^2+2 \Delta _{11}(r)-2}{2-r^2+2 \Delta _{11}(r)}.
\endeq
Then, one finds that the CF of the regular tetrahedron with unit sides is
\subsubsection*{2.1 Tetrahedron CF}
\begin{eqnarray} \gamma_{_{\rm T,a}}(r)&\equiv&
1-3\,\sqrt{\frac{3}{2}}\,\,r + \frac{3\left(2^{3/2}+\pi-\alft
\right)\,r^2}{\pi}-
\frac{\left(6+5\, (3)^{1/2}\, \pi \right)\, r^3}{2^{5/2}\,\pi },\label{Ta}
\end{eqnarray}
\begin{eqnarray}
\gamma_{_{\rm T,b}}(r)&\equiv&
\frac{3}{2^{5/2}\, r}  - 2   -  \frac{3\,\left((3)^{1/2}-3
\right)\, r}{2^{1/2}}
+\frac{3\left(2^{3/2}-\alpha _T-\pi \right)\, r^2 }{\pi }  -
\nonumber \\
& & \frac{\left(6-12\, \pi +5\, (3)^{1/2}\, \pi \right)\, r^3}
{2^{5/2}\, \pi },\label{Tb}
\end{eqnarray}
\begin{eqnarray}
 \gamma_{_{\rm T,c}}(r)&\equiv&
\frac{9+8\,(3)^{1/2}}{12\, (2)^{1/2}\,\, r}-6 +\frac{3\left(3+(3)^{1/2}
\right)\, r}{2^{1/2}}
+\frac{3\, \left(2^{3/2}-\alpha _T-3 \pi \right)\, r^2  }
{\pi }+\nonumber \\
& &
\frac{\left(12\,\pi-6+(3)^{1/2}\, \pi \right)\, r^3}
{2^{5/2}\, \pi },\label{Tc}
\end{eqnarray}
\begin{eqnarray}
\gamma_{_{\rm T,d}}(r) & \equiv& \frac{9+8(3)^{1/2}}
{24\, (2)^{1/2}\, r}+3
+\frac{9 \left(4+(3)^{1/2}\right) r}{2^{5/2}}+
\frac{3\, r^2\left(2^{3/2}-\pi-\alpha_{T}\right)}{\pi}-
\frac{\left(3-12 \pi +(3)^{1/2}\, \pi \right)\,r^3} {2^{3/2}\,\pi }-
                                            \nonumber \\
& &\frac{21\,r\,\Delta_{34}(r)}{2^{3/2}\,\pi}+
\frac{9}{12\, (2)^{1/2}\, \pi\,  r}\bigg[\arctan\left(\cT_1(r)\right)
-\frac{8\,(3)^{1/2}}{9} \arctan\left(\cT_2(r)\right)\bigg]+
                                         \nonumber\\
&& \frac{3}{\pi }\left[ \arctan\left(\cT_3(r)\right)
-3\,\arctan\left(\cT_4(r)\right)
-4\,\arctan\left(\frac{\cT_4(r)}{2}\right)+\frac{1}{2}\, \arctan
\left(\cT_5(r)\right)\right]          \nonumber\\
& &
-\frac{3\, r}{(2)^{1/2}\, \pi }\bigg[10\,\arctan\left(
\Delta_{34}(r)\right)
+5\,(3)^{1/2}\,\arctan\left((3)^{1/2} \Delta_{34}(r)\right)+
\arctan\left(\cT_5(r)\right)-     \nonumber \\
& &
6\,(3)^{1/2}\,\arctan\left(\cT_6(r)\right)+\frac{3^{1/2} } {2}
\arctan\left(\cT_{7}(r)\right) \bigg]+
                                                             \nonumber \\
& &
\frac{6 r^2}{\pi }\bigg[ \arctan\left(\cT_8(r) \right)+
\arctan\left(\cT_9(r)\right)\bigg]  -
\frac{3\, r^3}{2 \pi }\sqrt{\frac{3}{2}}\bigg[ \arctan
\left(\cT_{2}(r)\right)  -               \nonumber \\
& &
\arctan\left(\cT_6(r)\right)+
2\arctan\left(\cT_{10}(r)\right)-\frac{8}{3^{1/2}}\arctan
\left(\cT_{11}(r)\right)\bigg].
\end{eqnarray}
Here subscripts $a,\, b,\, c,$ and $d$ respectively refer to the
 $r$ intervals $[0,\,1/(2)^{1/2}]$,
$[1/(2)^{1/2},\, \sqrt{2/3}]$,  $[\sqrt{2/3},\,(3)^{1/2}/2]$  and
$[(3)^{1/2}/2,\,1]$ where the reported function definitions apply.
\subsubsection*{2.2 Octahedron CF}
The expression of the CF of the regular octahedron with unit edges reads
\begin{eqnarray}
\gamma_{_{\rm O,a}}(r)&\equiv &1-r\left(\frac{3}{2}\right)^{3/2} +
\frac{3 \left(\alpha _O-\pi +2 (2)^{1/2}\right)r^2 }{2\pi } -
\frac{\left(3-3 \pi +(3)^{1/2} \pi \right) r^3}{4 (2)^{1/2}\, \pi },
\quad\quad\quad\label{Oa}
\end{eqnarray}
\begin{eqnarray}
\gamma_{_{\rm O,b}}(r)&\equiv &\frac{2}{r}\,\sqrt{\frac{2}{3}}-3-
\frac{r}{2}\sqrt{\frac{3}{2}} +
\frac{3 r^2 \left(\alpha _O+\pi +2 (2)^{1/2}\right)}{2\pi }
-\frac{\left(3-3 \pi +7 (3)^{1/2}\, \pi \right)
   r^3}{4 (2)^{1/2}\, \pi },\quad\quad\label{Ob}
\end{eqnarray}
\begin{eqnarray}
\gamma_{_{\rm O,c}}(r)&\equiv &\frac{2}{r}\sqrt{\frac{2}{3}}-3-
\frac{r}{2} \sqrt{\frac{3}{2}}+
\frac{3 r^2 (\alpha _O+2 (2)^{1/2})}{2 \pi}-
\frac{\left(36(1- \pi) +5 (3)^{1/2}\, \pi\right )r^3}
{48 (2)^{1/2}\, \pi }-                     \nonumber\\
 & &
\frac{(17 r^2+3)\Delta_{34}(r)}{2 (2)^{1/2}\, \pi\,  r}  +
\frac{ r}{\pi }\sqrt{\frac{3}{2}} \,    \Big [
9\,\arctan\left(\cT_6(r)\right) + \arctan\left(3\ \cT_6(r)\right)\Big ]
+     \nonumber\\
 & &\frac{3\, r^2}{\pi }\arctan\left(\frac{T_4(r)}{2} \right) -
\frac{r^3}{8\, \sqrt{6}\, \pi} \Big[  18\, \arctan
\left(\cT_{6}(r)\right)  +     \label{OC}\\
&&
90\, \arctan\left(\frac{1}{3\,\cT_6(r)}\right)+24\,
\arctan\left(\cT_{7}(r)\right)-
8\, \arctan\left(\cT_{12}(r)\right)-   \nonumber\\
&&
\arctan\left(\cT_{13}(r)\right) -   6\, \arctan\left(\cT_{14}(r)\right)
 \Big],                     \nonumber
\end{eqnarray}
\begin{eqnarray}
\gamma_{_{\rm O,d}}(r)&\equiv &
\frac{4 (3)^{1/2}\,\pi-3 }{3 (2)^{1/2}\, \pi\,  r}+3 +
\frac{\left(4 (3)^{1/2}\, \pi-9\right) r}{3 (2)^{1/2}\, \pi }
+\frac{3 r^2}{4}
-\frac{\left(6-3 \pi +4 (3)^{1/2} \pi \right) r^3}
{8(2)^{1/2}\, \pi }+\nonumber\\
&&
\frac{(2)^{1/2}}{\pi\,r}\left(1+2\,r^2\right)\Delta_{11}(r)-
\frac{2\,\sqrt{6}}{\pi\,r}\arctan\left(T_{15}(r)\right)-
   \label{Od}\\
&&
\frac{12}{\pi}\arctan\left(T_{16}(r)\right)-\frac{2\sqrt{6}\,r}{\pi}
\arctan\left(T_{15}(r)\right)+
\frac{3\, r^2}{2 \pi}\arctan\left(\cT_{17}(r)\right)-\nonumber\\
 & &
\frac{r^3}{2\, (2)^{1/2}\, \pi} \left(2\, (3)^{1/2}\, \arctan
\left(\cT_{18}(r)\right)+
   3\, \arctan\left(\cT_{19}(r)\right)\right).\nonumber
\end{eqnarray}
Subscripts subscripts $a,\, b,\, c,$ and $d$  refer now  to
the r-intervals $[0,\,\sqrt{2/3}]$,
$[\sqrt{2/3},\, (3)^{1/2}/2]$,  $[(3)^{1/2}/2,\,1]$  and
$[1,\,(2)^{1/2}]$.
\subsubsection*{3 Discussion}
It has been checked that the reported CF expressions obey
the known constraints,
\ie\ $\gamma(0)=1$, $\gamma\p(0)=-S/4V$,
$\gamma(D_{max})=\gamma\p(D_{max})=0$, \quad $4\pi\times$
$\int_0^{D_{max}}r^2\gamma(r)dr=V$ and
$4\pi\int_0^{D_{max}}r^4\gamma(r)dr=2{R_{G}}^2V$,
$R_{\rm G}$ and $V$ denoting the Guinier giration radius and
the volume of the particle. \\
A graphical comparison of the behaviours of the  CFs of  the first three
Platonic solids,  the sphere and the cylinder (of height equal to its diameter)
is shown in Figure 1a.  Using the property that $\gamma_{D_{max}}(r)$,
 the CF of  a given shape particle  of maximal chord $D_{max}$,
is related to $\gamma(r)$, the CF of the particle with the same shape
and unit maximal chord, as  $\gamma_{D_{max}}(r)=\gamma(r/D_{max})$,
Figure 1 considers the case $D_{max}=1$ for all the particles, so that
the edge lengths of the tetrahedron, octahedron and cube are respectively 
equal to  $1,\ 2^{-1/2}$ and $3^{-1/2}$ while the sphere radius
is $1/2$ and the cylinder height  $2^{-1/2}$. It is noted that the choice
of comparing particles of equal maximal chord is more reasonable
than that of comparing particles of equal volume (see, \eg, Li \etal, 2011)
because the CF supports are equal.  Fig. 1a shows that the
CFs have a similar behaviour  and that, passing from the tetrahedron
to the octahedron, 
the cube and the cylinder, they approach that of the sphere. As
$r$ approaches to one, the
order is partly modified because the CF of the octahedron is smaller
than the tetrahedron's (a
feature not fully evident in the figure).
This property is related to the number of vertex pairs that are distant
$D_{max}$: this number
is 6 for the octahedron and 12 for the tetrahedron. 
  Figs. 1b and 1c respectively plot
the first and second derivative of the CFs considered in Fig. 1a.
Passing from Fig.1a to Fig.1c, one sees
that,  as the derivative order $n$ increases, the behaviours of
the $\gamma^{(n)}(r)$s
become more and more different in the five cases. One should in
particular observe the first
order discontinuities, located at  $r=1/(3)^{1/2}$ in the octahedron's
and cube's case and at
$r=1$ in the sphere's one, as well as the (infinitely high) cuspidal
behaviour located at
$r=1/(2)^{1/2}$ for the cylinder's case. The finite discontinuities are
generated by the
elliptic parallelism (Ciccariello, 1985) of the polyhedron faces and
the logarithmic one   by the hyperbolic
parallelism of the lateral surface of the cylinder (Ciccariello, 1989). 
The five intensities and their Porod plots are respectively shown in
Fig. 2a and 2b.
Note that in Fig. 2a the intensities have been scaled so as to make them
equal  at $q=0$.
Fig. 2a shows that, as the face number of the polyhedra increases, 
the intensities approach that of the sphere.  Fig. 2b, which refers to 
the non scaled intensities, reflects more details on the particle shapes, 
as  the highest $S/V$ value of the tetrahedron, the presence of oscillations 
whose spacing is $2\pi/d$ with $d$ equal to the
distance between the parallel facets. In the tetrahedron case
no parallelism is present and, consequently,
the small oscillations fade away as $q$ increases. [In fact,   they are
related to the first sub-asymptotic term decreasing as $q^{-9/2}$ as
it was shown  by  Ciccariello (2005b).]
Finally Fig. 3 shows the effect of a Poisson  size dispersion (Feigin \& Svergun,
1987), \ie\ $p_{4,1}(d)=d^4\,e^{-d}/4!$. The
intensities, smeared by this  size dispersion, are still different even though 
their differences are reduced and the oscillations are completely
washed out. \\ 
\subsection*{Acknowledgment}
I am grateful to Dr. Wilfried Gille for his critical reading of the manuscript. 
\vfill\eject
\section*{References}
\begin{description}
\item[\refup{}] Burger C. \&  Ruland, W. (2001). {\em Acta Cryst. A}
{\bf 57}, 482-490.
\item[\refup{}] Caccioppoli, R. (1956). {\em Lezioni di Analisi
Matematica}, (Treves, Napoli, vol.II).
\item[\refup{}]  Chiu, S. N., Stoyan, D.,  Kendall, W.S. \&
Mecke, J. (2013). {\em Stochastic Geometry and its Applications}, (
Wiley, Chichester, 3rd Ed.).
\item[\refup{}] Ciccariello, S. (2014).  {\em J. Appl. Cryst.} in the press. 
\item[\refup{}] Ciccariello, S. (2009). {\em J. Math. Phys.}
{\bf 50}, 103527(1-10).
\item[\refup{}] Ciccariello, S.  (2005b).  {\em Fibers \& Textiles in East. Eur.}
{\bf 13}, 41-46.
\item[\refup{}] Ciccariello, S.  (2005a).  {\em J. Appl. Cryst.}
{\bf 38}, 97-106.
\item[\refup{}] Ciccariello, S.  (1989).  {\em Acta Cryst. A}
{\bf 45}, 86-99.
\item[\refup{}] Ciccariello, S.  (1985).  {\em Acta Cryst. A}
{\bf 41}, 560-568.
\item[\refup{}] Debye, P.  \&   Bueche, A.M. (1949). {\em J. Appl. Phys.}
{\bf 20}, 518-525.
\item[\refup{}] Feigin, L.A. \&    Svergun, D.I. (1987). {\em Structure Analysis
by Small-Angle X-Ray \&  Neutron Scattering}, (Plenum Press, New York).
\item[\refup{}]   Gille, W. (2013). {\em Particle and Particle Systems
characterization }, (CRC Press, London)
\item[\refup{}] Gille, W. (1999). {\em J. Appl. Cryst.}  {\bf 32}, 1100-1105.
\item[\refup{}] Gille, W.  (1987).  {\em Exp. Tech. Phys.}  {\bf 35}, 93-99.
\item[\refup{}] Goodisman, J. (1980).  {\em J. Appl. Cryst.}  {\bf 13}, 132-135.
\item[\refup{}] Guinier, A.  \&  Fournet, G.    (1955). {\em Small-Angle
Scattering of X-rays}, (Wiley, New York).
\item[\refup{}] Li, X., Chewn, C. -Y., He, L., Meilleur, F., Myles, D.A.A.,
Liu, E., Zhang, Y., Smith, G.S., Herwig, K.W., Pynn, R. \& Chen, W.-R.
(2011).  {\em J. Appl. Cryst.} {\bf 44}, 345-557.
\end{description}
\vfill\eject                
\begin{figure}[!h]
{
\includegraphics[width=7.truecm]{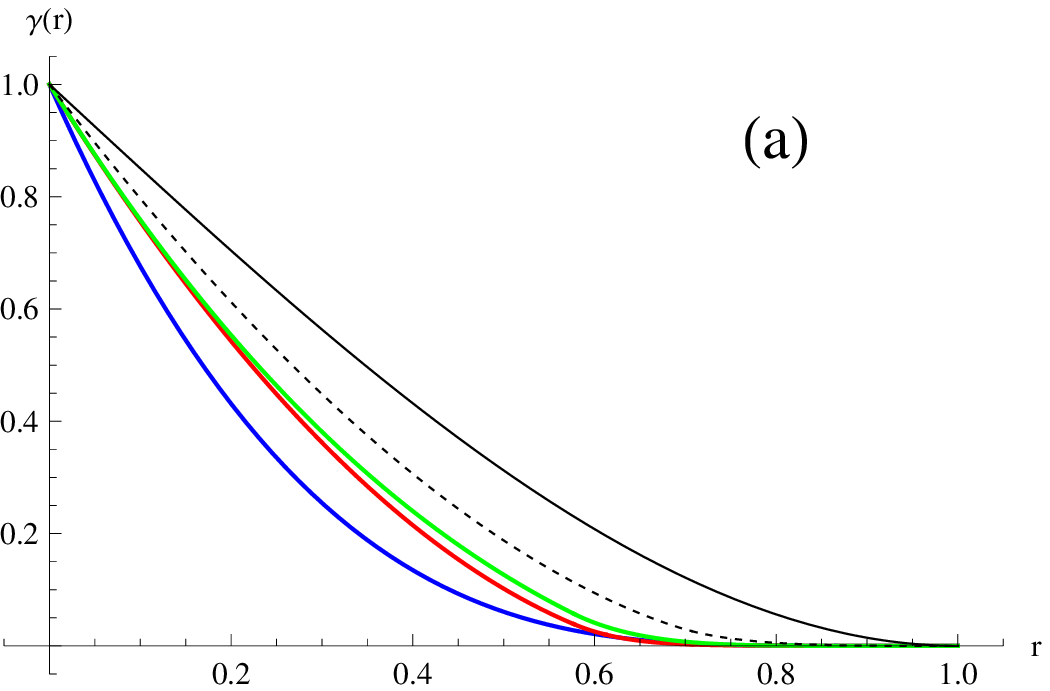}
\includegraphics[width=7.truecm]{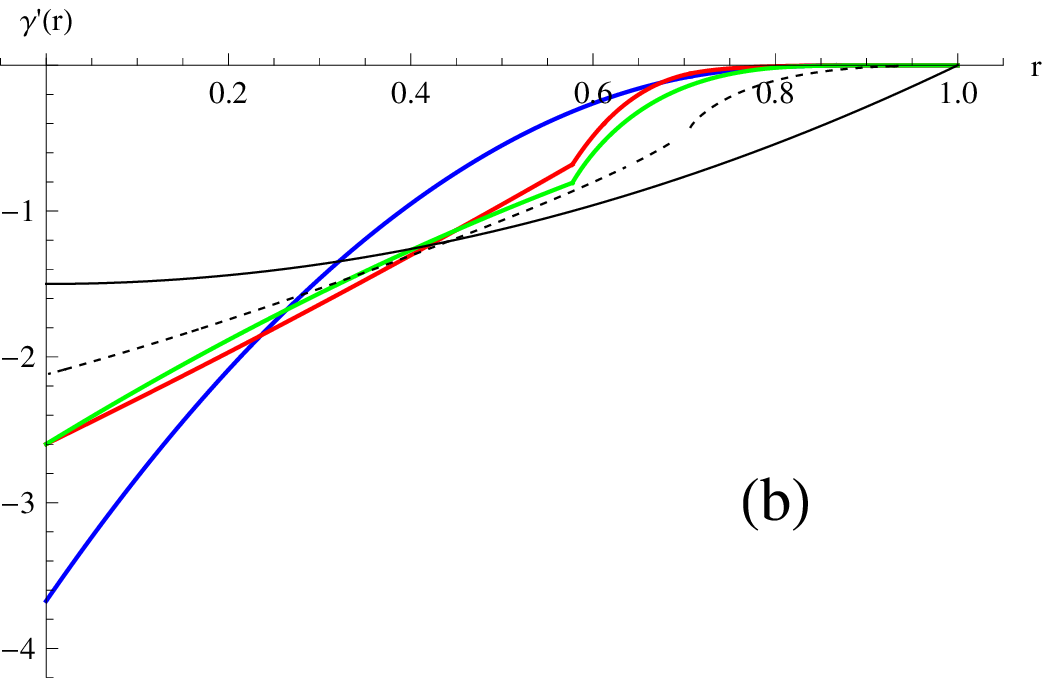}
\includegraphics[width=7.truecm]{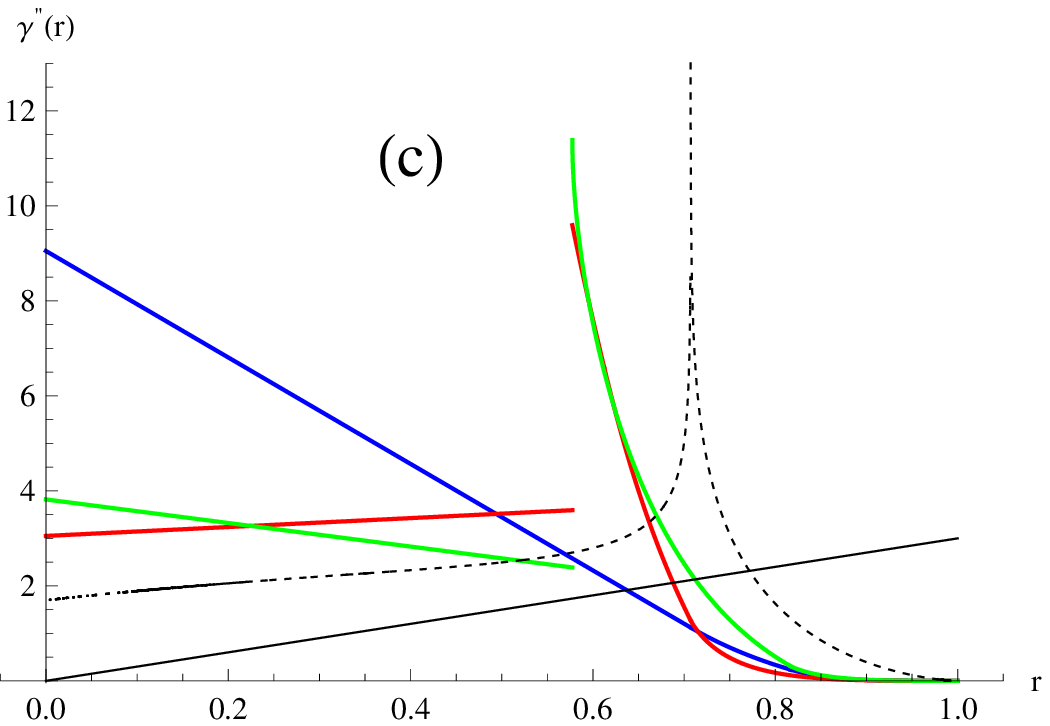}
}
\caption{\label{Fig1} {{(a): Behaviour of the CFs of the tetrahedron
(thick blue curve), octahedron (thick red), cube (thick green), cylinder
(thin dotted black) and sphere (thin black). [All these solids have their
largest diameter $D_{max}$, named $L$ in Figures 2 and 3,  equal to one. 
Further, the  height and the diameter of the cylinder are equal]; 
(b) behaviour of the first derivative of the CFs;
(c): behaviour of the second derivative of the CFs  }}}
\end{figure}
\vfill\eject
\begin{figure}[!h]
{
\includegraphics[width=7.truecm]{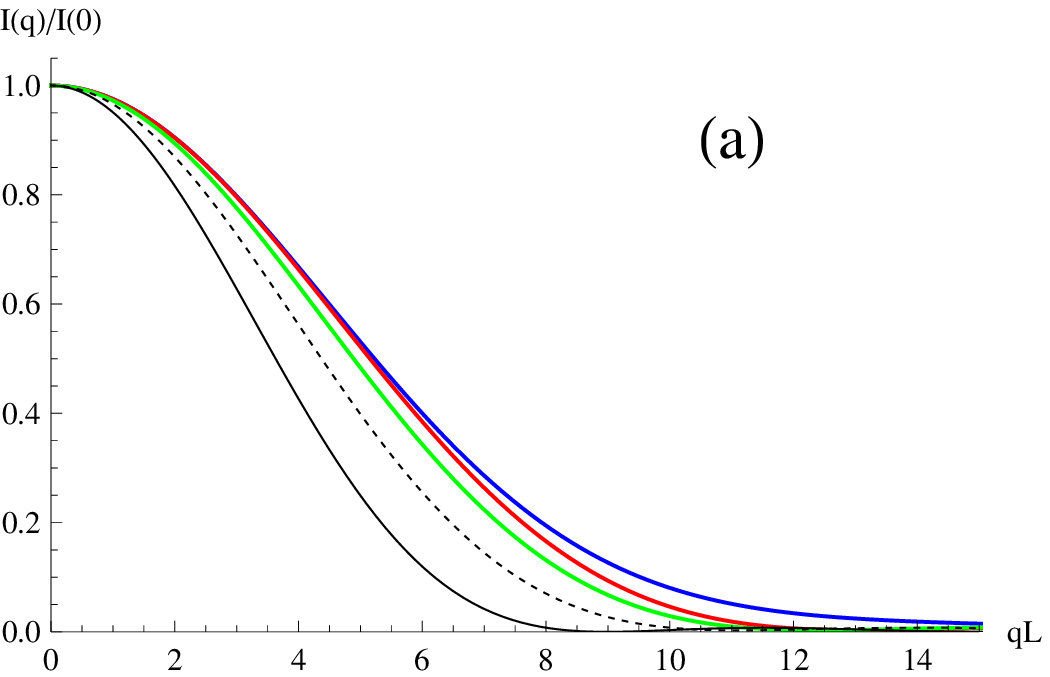}
\includegraphics[width=7.truecm]{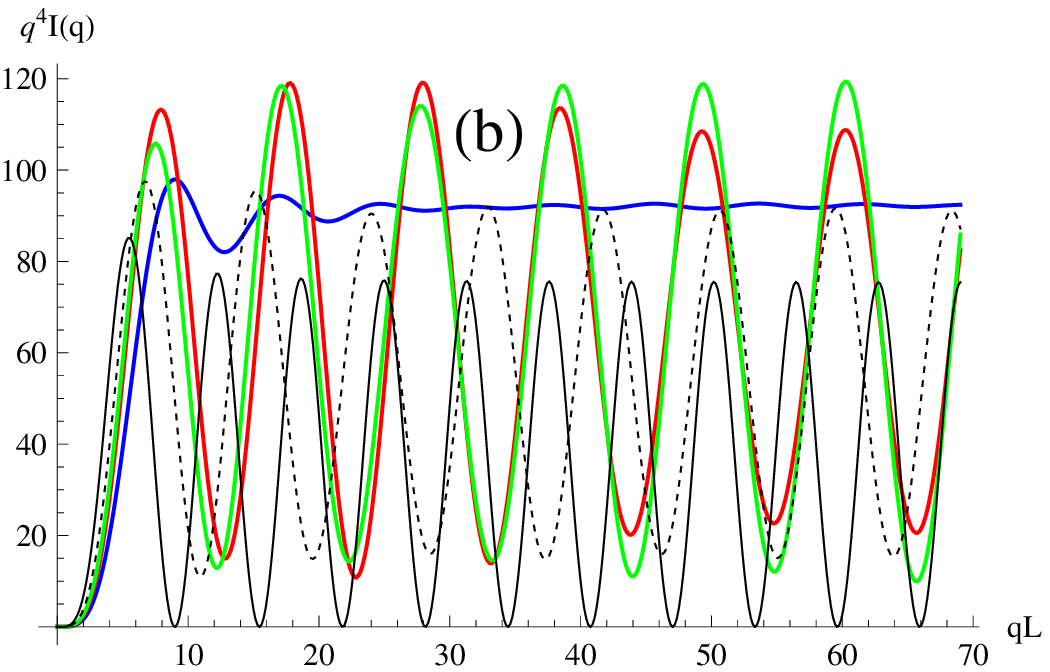}
}
\caption{\label{Fig2}{{  (a): Behaviour of the scattering intensities  relevant
to the solids  considered in Fig.1 with  the same colour meaning. Note that
the intensities have been scaled to 1 at q=0;  (b): Porod
plot of the previous non-scaled intensities. [$L$ is equal to $D_{max}$.] }}}
\end{figure}
\vskip 2.truecm
\begin{figure}[!ht]
{
\includegraphics[width=7.truecm]{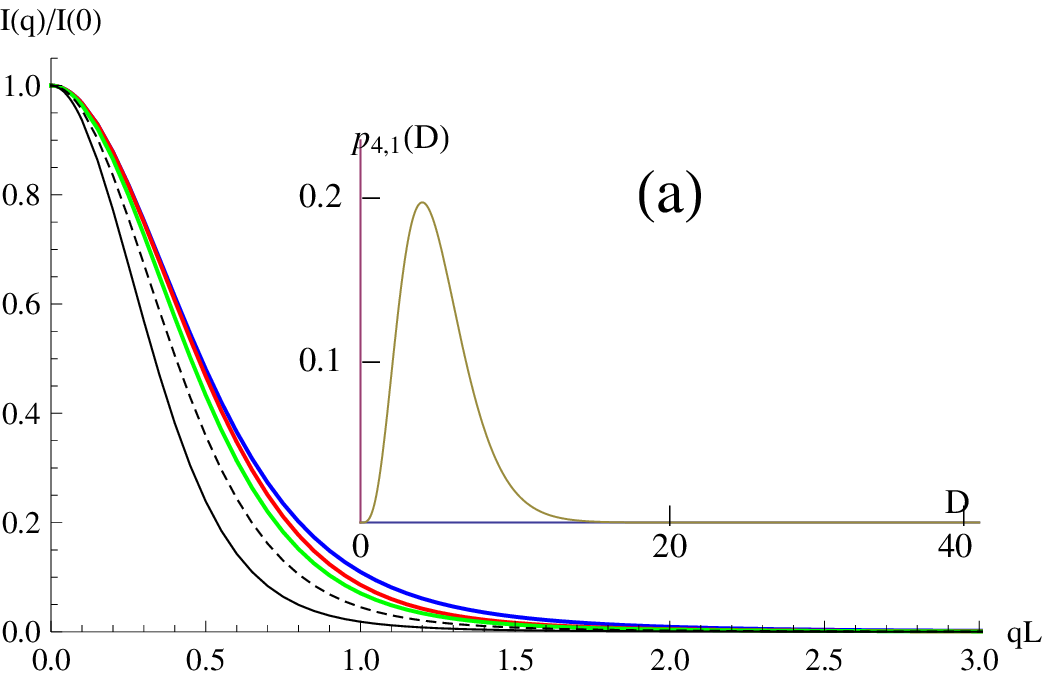}
\includegraphics[width=7.truecm]{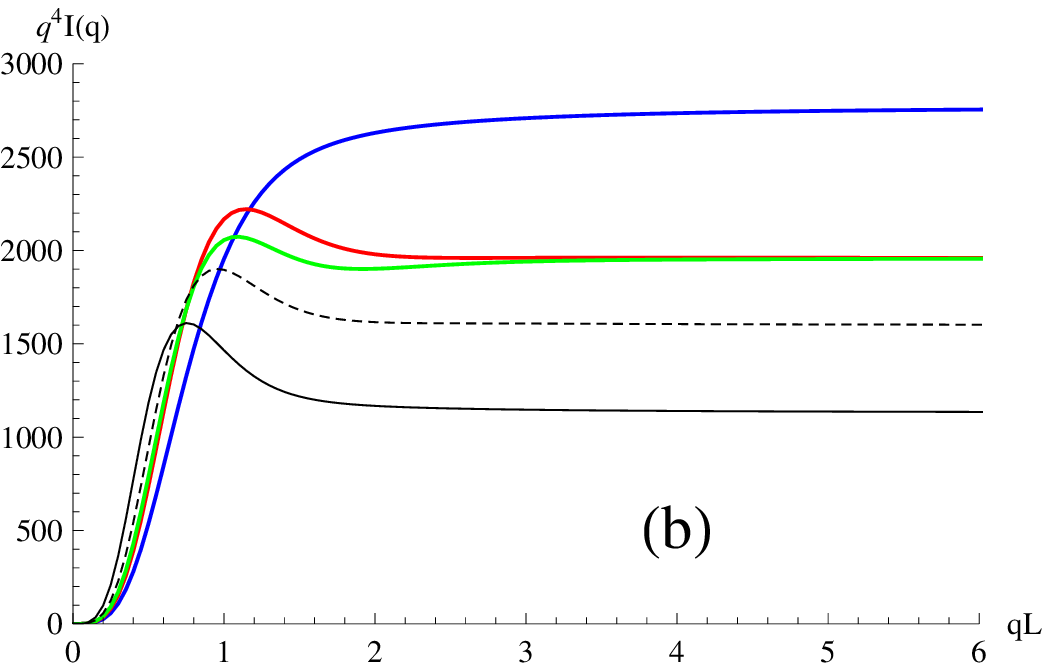}

}
\caption{\label{Fig3}{{(a): Behaviour of the scattering intensities (scaled as in Fig.2a)
relevant to polydisperse distributions of the particles considered in Fig.1. The particle
size probability density  is the Poisson one with $n=4$ and $\lambda=1$,
\ie\ $p_{4,1}(d)=d^4\, e^{- d}/4!$ and is shown in the inset; (b) Porod plot
of the polydisperse non-scaled intensities.  }}}\end{figure}
\end{document}